# Starlink Satellite Brightness Before VisorSat

Anthony Mallama

14012 Lancaster, Bowie, MD, 20715, USA

anthony.mallama@gmail.com

2020 June 15

**Abstract**

The mean of 830 visual magnitudes adjusted to a distance of 550 km (the operational altitude) is 4.63 +/-0.02. The data on DarkSat, the low-albedo satellite, indicate that it is fainter than the others by 1.6 magnitudes or 78%. However, there is considerable uncertainty in this value due to the small number of observations. Some satellites were observed to flare by 10,000 times their normal brightness. These statistics can serve as a baseline for assessing the reduced brightness of the VisorSat design for future Starlink satellites.

# 1. Introduction

The many artificial satellites orbiting the Earth are beginning to interfere with ground-based astronomical observing programs. Those in low-earth-orbit (LEO) are a special concern because of their great apparent brightness. Thousands more LEO satellites are currently being launched by the SpaceX company. The impact of these Starlink satellites on astronomy has been studied by Gallozzi et al., 2020, Hainaut and Williams, 2020, and McDowell, 2020. While those papers discuss many aspects of the satellites their brightness is only addressed rather generally. This paper provides a comprehensive analysis and summary of Starlink magnitudes through May 2020. SpaceX has designed a new model of Starlink satellites, called VisorSat, which includes a Sun shade to make it appear dimmer in the sky. The results from this paper can serve as a reference to compare with the brightness of VisorSat.

Section 2 discuses three factors that affect Starlink satellite brightness. The first is distance where brightness varies in accordance with its inverse square. The second factor is the phase of the satellite's illumination by the Sun, as more fully illuminated satellites are expected to be brighter. Third is the orientation of the flat-panel which forms the body of satellites. Starlink satellites are brighter when this panel is more perpendicular to the Sun and to the observer.

Section 3 describes the sets of Starlink magnitudes collected by an automated observatory and visual observers. Section 4 reports on the fits of these observed magnitudes to models for phase angle and flat-panel orientation. A characteristic Starlink magnitude is determined and the variance around that value is discussed.

Some extremely bright 'flaring' of Starlink satellites has been reported. These events can produce a brightness surge of 10,000 times or more for a brief interval. Section 5 describes the celestial geometry associated with this phenomenon and explains its probable cause.

Section 6 focuses on the special Starlink satellite known as DarkSat. SpaceX applied a low-albedo coating to DarkSat as a test. The darkening was meant to address the concerns of astronomers regarding satellite brightness. The available observations of DarkSat are analyzed and the effectiveness of the low-albedo coating is assessed.

Section 7 briefly describes the design of VisorSat which includes a sun shade to reduce brightness. Additional steps being taken by SpaceX to reduce the brightness of Starlink satellites are also mentioned.

## 2. Brightness Models

Satellite magnitudes vary in accordance with the inverse-square law of light like any other celestial body. Distance is taken to be that between the satellite and the observer. The distance between the satellite and the Sun is ignored because it only varies by a few percent and, thus, has very little effect on brightness. All the magnitudes analyzed in this paper were first reduced to the standard satellite distance of 1000 km.

The brightness of any object illuminated by the Sun may also depend on the phase of illumination The parameter associated with this dependency is the phase angle, which is defined as the arc between the Sun and the observer as measured at the body. As an example of its use, Mallama et al. (2017) employ phase angle to characterize the magnitudes of the planets of the solar system.

There are two common forms of the phase angle model, geometrical and empirical. The geometrical model calculates the illuminated fraction of a spherical body as a function phase angle and determines brightness accordingly. This analytical approach does not apply very well to Starlink satellites because they are in the shape of a flat panel.

The empirical phase angle model is used in this paper because it is suited to a body of any shape. The mathematical representation is the best fitting slope of magnitude versus phase as determined from the observations. Thus, slope and absolute magnitude are the two parameters to fit in the phase angle model.

Since the Starlink satellites are shaped like flat panels, brightness also depends on their orientation in space relative to the Sun and observer. Mallama (2020) derived a mathematical representation customized to the shape and orientation of Starlink satellites. Practical implementation of this model depends upon the assumption that the two broad surfaces of the flat panel face nadir and zenith, which is the approximate orientation expected when the satellites reach their operational altitude. Brightness is represented by the product of the projections of the nadir-facing side to the observer and to the Sun as illustrated in Figures 1 and 2.

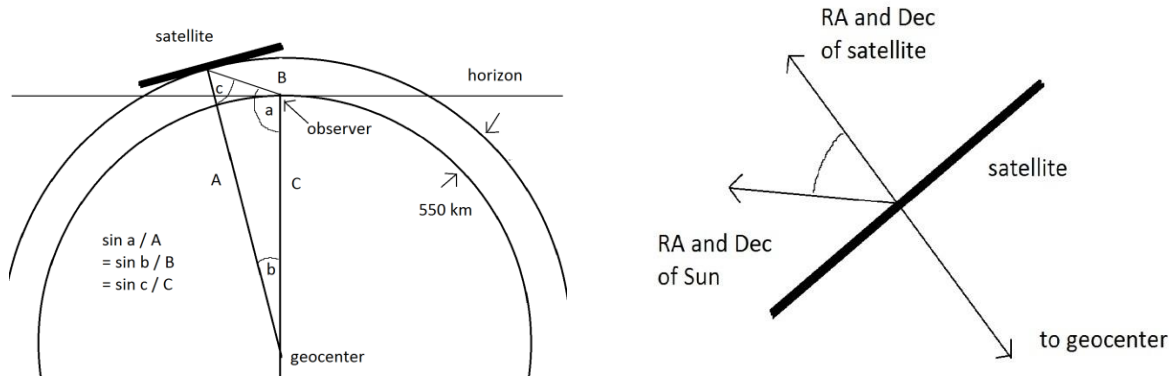

Figure 1 (left) illustrates the observer aspect of the flat panel model. The angle between the perpendicular to the flat panel and the direction to the observer is labeled *c*. The cosine of this angle (*M* in equation 1 of this paper) is the projection of the panel to the observer.

Figure 2 (right) shows the Sun aspect of the flat panel model. This is the angle between the perpendicular to the flat panel and the direction to the Sun. The Illumination of the nadir-pointing side of the satellite is the supplementary angle to that shown in the figure and its cosine is *N* in equation 1.

## 3. Observations

Several visual observers and one automated observatory have recorded magnitudes of Starlink satellites. These sources are listed in Table 1 along with their geographic coordinates. Generally, the visual observers used the variable star technique where the magnitude of an object is estimated by comparison to one or more reference stars. The magnitudes of the references were taken from Tycho 2 and similar photometric catalogs. The data were retrieved from the SeeSat-L mail archive hosted at http://satobs.org.

Many additional magnitudes were taken from the on-line database of the Mini-MegaTORTORA (MMT-9) automated observatory (Karpov et al. 2015). This instrument operates at visible wavelengths and the observations used in this study were acquired in the 'clear' filter. S. Karpov (private communication) indicates that the MMT-9 magnitudes are close to the V band-pass for objects with small B-V color indices and, in fact, they are found to be consistent with those of the visual observers. The data were downloaded from http://mmt9.ru/satellites/.

The observations are divided into three groups for later analysis. The largest group contains magnitudes of Starlink satellites at their 550 km operational altitude. The other groups are for observations of brightly flaring satellites and for observations of DarkSat.

Table 1. Observations

```
                                <---- Observations ---->
Name           Latitude Longitude   550-km  Flares  DarkSat
A. Amorim       27.7 S    48.5 W      16       4       1
R.E. Cole       51.1 N     0.5 W       0       0       1
R. Eberst       55.9 N     3.1 E      76       0       0
A. Mallama      39.0 N    76.8 W      21       0       1
T. Molczan      43.7 N    79.4 W       0       0       1
J. Respler      40.3 N    74.4 W      65       0       2
B. Young        36.1 N    96.0 W      98       4       0
MMT-9           43.6 N    41.4 E     554       0       0
Total                                830       8       6
```

T. Molczan (private communication) prepared all the 550 km magnitudes that are analyzed in the next section. This included collecting the observations, adjusting observed magnitudes to the 1000 km distance, computing the phase angles and calculating the two angles needed for flat-panel modeling.

### 4. Satellites at the operational altitude

This section reports on analysis of the data of Starlink satellites at their 550 km operational altitude. The mean magnitude at the standard distance of 1000 km is computed and a 'characteristic' magnitude corresponding to the operational altitude is calculated. Then the scatter in the magnitudes is discussed. Finally the data are fit to the flat-panel model and to the empirical phase function.

### 4.1 Mean and characteristic magnitudes

The average magnitude adjusted to the standard distance of 1000 km for the Starlink satellites observed at the operational altitude is 5.93. Adjustment of this value to 550 km the gives the 'characteristic' magnitude of 4.63. This magnitude characterizes the brightness of a Starlink satellite at its operational altitude when seen near zenith. The standard deviation of the 830 magnitudes and the standard deviation of the mean are 0.67 and 0.02 magnitude, respectively.

### 4.2 Variation

The standard deviation of 0.67 magnitudes has multiple components. The first of these is the uncertainty of the reported magnitudes. The visual observers employed a variety of measurement techniques. Some used multiple reference stars for each measurement while other used only a single star. One observer only reported magnitudes to the nearest integer number. Data from the MMT-9 also shows considerable scatter for a single satellite on a timescale of seconds.

Another component of the scatter appears to be related to satellite orientation. Several observers have recorded large magnitude differences for satellites in practically the same orbit observed close together in time. An example is this quote from an experienced observer "there were 4 pairs [of satellites] with second following first by several seconds. In each case the first was mag 4-6. The second was 2-3 mags brighter" (http://www.satobs.org/seesat/Mar-2020/0126.html). These satellites were at the operational altitude and differing orientations is a likely explanation for the large brightness variations.

A final component may be related to the observer's geographic location. The magnitudes from the observer located furthest north are fainter than the rest. This observer is also very experienced so a large systematic error seems unlikely. Orientation of the Starlink satellites relative to his location is a possible explanation.

### 4.3 Model Fits

After adjustment to a distance of 1000 km the observations were fit to the flat-panel model and the phase angle model described in section 2. The parameter of the flat-panel fit is the term

$$x_{FP} = -2.5 \log 10 \, ( M \, N )$$

equation 1

where M and N are illustrated in Figures 1 and 2, respectively. The term '-2.5 log 10' converts from brightness to magnitude.

The parameter of the phase angle fit, $x_{PA}$ is that angle measured in radians. Thus the equation of the combined fit is

$$y = m_{FP} x_{FP} + m_{PA} x_{PA} + b$$

equation 2

where *y* is the set of magnitudes, the *m* terms are the derived slopes for the flat-panel and phase angle models and *b* is a constant. The same equation was used to solve for the each of the two models separately by setting the terms of the other model to zero. Thus, for flat-panel and phase angle

$$y = m_{FP} x_{FP} + b$$

equation 3

and

$$y = m_{PA} x_{PA} + b$$

equation 4

respectively.

The statistical results for the three fits are listed in Table 2. For the combined fit the slope for the flat-panel model was highly significant with a 7 sigma detection while that of the phase angle fit was only marginally significant at 2 sigma. For the individual fits the slope of the flat-panel model was again significant at 7 sigma while that for the phase angle model was not significant. The individual fits for flat-panel and phase angle are illustrated in Figures 3 and 4.

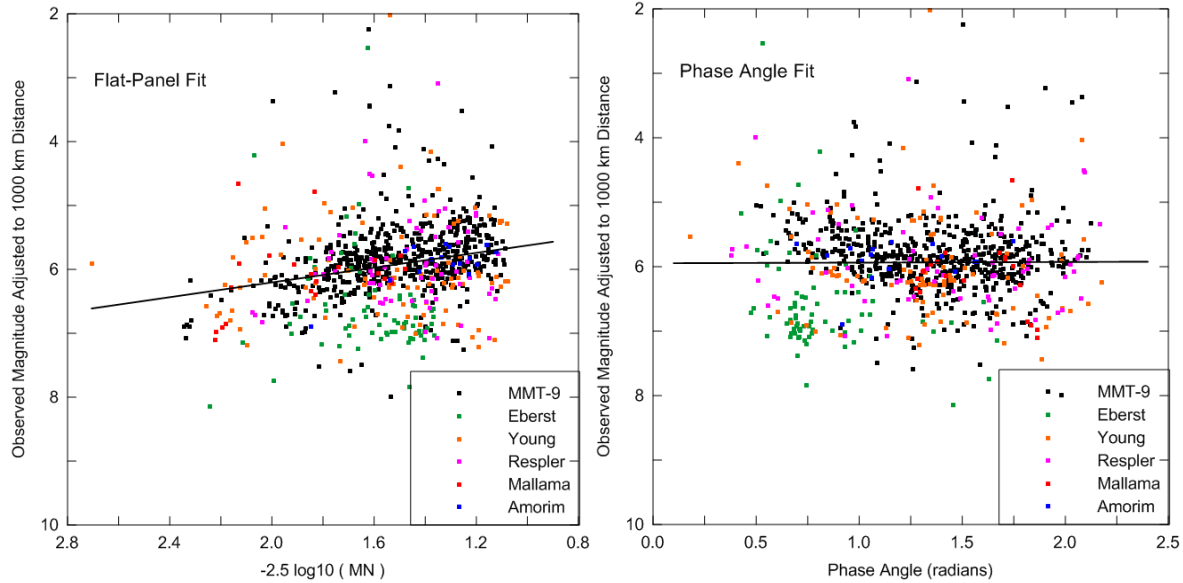

Figures 3 (left) and 4 (right) illustrate the fit of the flat-panel model and the phase angle model, respectively, to the observed magnitudes corrected to 1000 km distance.

Despite the significance of the slopes for these models the standard errors of the magnitudes were reduced very little. Table 2 shows that for no model at all the standard error is 0.672 while even the combined model only reduced it to 0.653.

Table 2. Model fits

```
                  mFP                mPA               b              r2       se
              ------------       ------------      ------------     -----    -----
              slope   +/-        slope   +/-       const   +/-
Combined      0.610   0.086     -0.100   0.059     5.129   0.140    0.057    0.653
Flat-panel    0.578   0.084                        5.046   0.131    0.054    0.654
Phase angle                     -0.010   0.059     5.946   0.081    0.000    0.672
No model                                                                     0.672
```

**5. Flare Events**

Multiple extremely bright flares of Starlink satellites were observed during the ten day period beginning on 2020 April 8. In some cases the satellites exceeded their normal brightness by more than 10,000 times (10 magnitudes). These satellites were between altitudes 380 and 425 km.

The celestial geometry of the brightest events is fairly consistent. The most striking aspect is that the azimuth of the satellite is approximately equal to that of the Sun in most cases. The root-mean-square of the azimuth differences was 28 degrees.

This geometry suggests that the flares are due to specular reflection of sunlight from the nadir-facing side of these flat-panel shaped satellites. Pictures on the SpaceX web site *Starlink.com* appear to show several such reflective areas on the nadir side of the spacecraft body.

Table 3. Bright Flare Observations

| Observer | Starlink # | UTC Date | UTC Time | Magnitude |
|---|---|---|---|---|
| Amorim | 1288 | 2020-04-08 | 22:40:10 | -6 |
| Amorim | 1255 | 2020-04-10 | 21:37:45 | -7 |
| Amorim | 1277 | 2020-04-10 | 22:13:17 | -3 |
| Amorim | 1262 | 2020-04-10 | 22:19:47 | -3.5 |
| Young | 1258 | 2020-04-17 | 02:27:08 | -4 |
| Young | 1296 | 2020-04-17 | 02:27:22 | -8 |
| Young | 1292 | 2020-04-17 | 02:27:50 | -8 |
| Young | 1260 | 2020-04-17 | 02:28:22 | -4 |

Another more detailed hypothesis which seeks to explain Starlink flares was proposed by R.E. Cole (http://www.satobs.org/seesat/May-2020/0062.html). His model emphasizes the roll angle of the satellites and its relationship to the solar direction.

A third explanation has been proposed by R. Herrick (http://www.satobs.org/seesat/apr-2020/0252.html). He has stated that the flare events are of very brief duration and suggests that the motion of the satellite in elevation (that is, in the Sun-satellite plane) accounts for the flares.

**6. DarkSat**

SpaceX applied a low-abledo coating to one satellite, called DarkSat, in order to reduce its brightness. This was a first attempt to make the Starlink satellites less problematic for astronomical researchers. One electronic measurement and several visual estimates have been obtained since DarkSat reached its operational altitude.

Tregloan-Reed et al. (2020) observed Dark Sat and a comparison Starlink satellite in the g' band-pass of the Sloan photometric system. They determined that the coating reduced its reflectivity by 55%. While that electronic result is very accurate, considerable uncertainty remains about the actual reduction of brightness. There is large scatter in Starlink magnitudes and it is probably attributable to their orientation relative to the Sun and observer as noted in section 4.2. Therefore, we examine the 6 visual observations that are available.

Table 4 lists the difference between the observed magnitudes of DarkSat adjusted to 1000 km relative to the mean magnitude, 5.93, determined for other Starlink satellites in section 4.1. Five of the 6 values are positive indicating that DarkSat is fainter than the other satellites. The average of these positive values is 1.6 magnitudes, corresponding to a 78% brightness reduction, which is somewhat more than the electronic result.

However, the first observation in the table indicates that DarkSat was about 3 magnitudes *brighter* than the average of the other satellites. This observation was made by an experienced observer and it serves to emphasize the uncertainty in the actual reduction in the effectiveness of the low-albedo coating on DarkSat. In any case, the DarkSat design is being abandoned by SpaceX in favor of that described in the following section.

Table 4. DarkSat Magnitudes

| Observer | UTC Date | Delta Magnitude |
|---|---|---|
| J. Respler | 2020-02-15 | -3.0 |
| R. Cole | 2020-03-01 | +1.2 |
| T. Molczan | 2020-04-03 | +2.5 |
| A. Amorim | 2020-04-16 | +0.9 |

```
J. Respler      2020-05-02      +2.0
A. Mallama      2020-05-31      +1.5
```

**7. VisorSat**

SpaceX is taking a proactive approach in working with the astronomical community to reduce the brightness of Starlink satellites (https://spacenews.com/spacex-claims-some-success-in-darkening-starlink-satellites/). Since the low-albedo coating on DarkSat was not entirely effective in reducing brightness to levels where it will not significantly interfere with astronomical observations, the company is implementing a different approach for future satellites. SpaceX has designed a 'visor' which will shade the reflective underside of the satellite body from sunlight and thus render it dark. They are also making software changes so that the satellites will be more edge-on to the Sun during certain phases of their lifetimes. We do not discuss these changes in detail here but point out that the results in section 4 of this paper can be used to assess the effectiveness of darkening the VisorSat satellites after observations of that model in space have been obtained.

**8. Conclusion**

This paper summarizes the brightness of Starlink satellites prior to the VisorSat model. Their mean observed magnitude, the brightness of their flares and the reduced brightness of DarkSat are reported.

The mean of 830 observed magnitudes of Starlink satellites at the operational altitude adjusted to the standard satellite distance of 1000 km is 5.93. Adjustment of this value to 550 km the gives the characteristic magnitude of 4.63. This is the approximate brightness of a Starlink satellite at the operational altitude when observed near zenith. The standard deviation of the mean is 0.02 magnitude.

The standard deviation of the observations around the mean is 0.67 magnitude. The scatter is partly due to observational uncertainty and partly due to variation of the orientation of the satellites relative to the Sun and to the observer.

The magnitudes were fit with a model for the shape and presumed nadir-facing orientation of the flat-panel shaped satellites. There is a very significant 7 sigma correlation between the flat-panel model and the data. However, the model does not greatly reduce the scatter. A fit between the magnitudes and a phase angle model is only marginally significant at 2 sigma and does not reduce the scatter appreciably.

Extremely bright flares of Starlink satellites were observed in 2020 April. The observed brightness of the satellites exceeded their normal brightness by more than 10,000 times in some cases. The celestial geometry of the brightest events suggests that flaring is caused by the specular reflection of sunlight from the nadir facing side of the satellites.

Visual observations of DarkSat, the low-albedo satellite, suggest that it is 1.6 magnitude or 78% dimmer than other Starlink satellites. However, one discordant observation found it to be brighter than the others. This anomaly emphasizes the variability of these satellites.

The new design for Starlink satellites called VisorSat is briefly discussed. It is hoped that this design will greatly reduce the interference of these satellites with astronomical observations.

**Acknowledgments**

Communications with Richard E. Cole regarding the flaring of Starlink satellites and concerning their surface characteristics were very valuable. David R. Skillman provided several insights on the shape and structure of the satellites. He also reviewed and commented on an earlier version of the manuscript. Ted Molczan offered some useful ideas for the data analysis.

**References**

Cole, R.E. 2020. Measurement Of The Brightness Of The Starlink Spacecraft Named "DARKSAT" Research Notes of the AAS. 4, 3. https://iopscience.iop.org/article/10.3847/2515-5172/ab8234

Gallozzi, S., Scardia, M., and Maris, M. 2020. Concerns about ground based astronomical observations: a step to safeguard the astronomical sky. https://arxiv.org/pdf/2001.10952.pdf.

Hainaut, O.R., and Williams, A.P. 2020. Impact of satellite constellations on astronomical observations with ESO telescopes in the visible and infrared domains. *Astron. Astrophys.* manuscript no. SatConst. https://arxiv.org/abs/2003.019pdf.

Karpov, S., Katkova, E., Beskin, G., Biryukov, A., Bondar, S., Davydov, E., Perkov, A. and Sasyuk, V. 2015. Massive photometry of low-altitude artificial satellites on minimegaTORTORA. Fourth Workshop on Robotic Autonomous Observatories. RevMexAA.

Mallama, A., Krobusek, B. and Pavlov, H. 2017. Comprehensive wide-band magnitudes and albedos for the planets, with applications to exo-planets and Planet Nine. *Icarus*, 282, 19-33. https://doi.org/10.1016/j.icarus.2016.09.023.

McDowell, J. 2020. The low Earth orbit satellite population and impacts of the SpaceX Starlink constellation. *ApJ Let*, 892, L36 and https://arxiv.org/abs/2003.07446

Tregloan-Reed, J, Otarola, A., Ortiz, E., Molina, V., Anais, J., Gonzalez, R., Colque, J.P. and Unda-Sanza, E. 2020. First observations and magnitude measurement of SpaceX's Darksat. *Astron. Astrophys*., manuscript no. Darksat_Letter_arXiv_submission_V2. https://arxiv.org/pdf/2003.07251.pdf.
13